\documentclass{aa}

\usepackage{txfonts}	
\usepackage{graphicx}
\usepackage{natbib}
\bibpunct{(}{)}{;}{a}{}{,} 

\begin{document}

\title{Quasar jet emission model applied to the microquasar GRS\,1915+105}

\author{M. T\"urler\inst{1}\fnmsep\inst{2}
	\and T. J.-L. Courvoisier\inst{1}\fnmsep\inst{2}
	\and S. Chaty\inst{3}\fnmsep\inst{4}
	\and Y. Fuchs\inst{4}}

\offprints{M. T\"urler,
	\email{Marc.Turler@obs.unige.ch}}

\institute{INTEGRAL Science Data Centre, ch. d'Ecogia 16, 1290 Versoix, Switzerland
	\and Observatoire de Gen\`eve, ch. des Maillettes 51, 1290 Sauverny, Switzerland
	\and Universit\'e Paris 7, 2 place Jussieu, F-75\,005 Paris, France
	\and Service d'Astrophysique, DSM/DAPNIA/SAp, CEA/Saclay, F-91\,191 Gif-sur-Yvette, Cedex, France}

\date{Received 18 December 2003 / Accepted 14 January 2004}

\abstract{
The true nature of the radio emitting material observed to be moving
relativistically in quasars and microquasars is still unclear. The microquasar
community usually interprets them as distinct clouds of plasma, while the
extragalactic community prefers a shock wave model. Here we show that the
synchrotron variability pattern of the microquasar GRS\,1915+105 observed on 15
May 1997 can be reproduced by the standard shock model for extragalactic jets,
which describes well the long-term behaviour of the quasar 3C\,273. This
strengthens the analogy between the two classes of objects and suggests that
the physics of relativistic jets is independent of the mass of the black hole.
The model parameters we derive for GRS\,1915+105 correspond to a rather
dissipative jet flow, which is only mildly relativistic with a speed of
$0.60\,c$. We can also estimate that the shock waves form in the jet at a
distance of about 1\,AU from the black hole.

}

\maketitle

\section{Introduction}
\label{sec:intro}

Microquasars are high-energy binary systems exhibiting jets with apparent
superluminal motion suggesting that they are the miniature replicates in our
Galaxy of the distant quasars \citep{MR98}. \object{GRS\,1915+105} was the
first microquasar discovered \citep{MR94} and can be considered as an
archetypical object of its class. Its radio emission can be divided into three
distinct states: a plateau jet, pre- and post-plateau flares and radio
oscillation events \citep{KFP02}. The flares are the most powerful events.
Their emission arises from two emitting blobs observed to move relativistically
in opposite directions up to about ~3000\,AU from the core \citep{FGM99}. This
has been interpreted as evidence for pairs of plasma clouds symmetrically
ejected on both sides of the accretion disc \citep{RM99}. This plasmoid
ejection model is also used to explain the less powerful oscillation events
arising in a shorter jet extending only up to $\sim$ 30--50\,AU \citep{DMR00}.
This interpretation is supported by the observation of simultaneous X-ray dips,
suggesting that material from the disappearing inner accretion disc is ejected
from the black hole vicinity \citep{BMK97a,BMK97b,MDC98,EMM98}.

Observational evidence for this phenomenon has recently also been obtained for
the active galaxy 3C\,120 \citep{MJG02}. This discovery, by strengthening the
analogy between galactic and extragalactic objects, rises the question on the
true nature of the synchrotron emitting material. Is the microquasar plasmoid
model or the quasar shock wave model the correct interpretation\,? \citet{AA99}
showed that expanding clouds of relativistic plasma cannot describe the
observations unless there is an in situ electron acceleration process. An
alternative model of a variable continuous jet also fails to explain the
observed flux oscillations in GRS\,1915+105 as it leads to unphysically large
electron densities \citep{CKC03}. There is therefore growing evidence that the
emitting electrons must be reaccelerated along the jet, as it is the case in a
shock wave mechanism \citep{KSS00}.

To further test the shock wave interpretation, we apply here to GRS\,1915+105
the same model used to describe the long-term variability behaviour of
\object{3C\,273} from submillimeter to radio wavelengths \citep{TCP00}. This
model derived from the model of \citet{MG85} describes analytically the
evolution of synchrotron emission resulting from a shock wave moving down a
relativistic jet. Compared to the model of \citet{V66} for an expanding plasma
cloud it has the advantage to include the effects of both synchrotron and
inverse-Compton energy losses.

\begin{figure*}
\includegraphics[width=12cm]{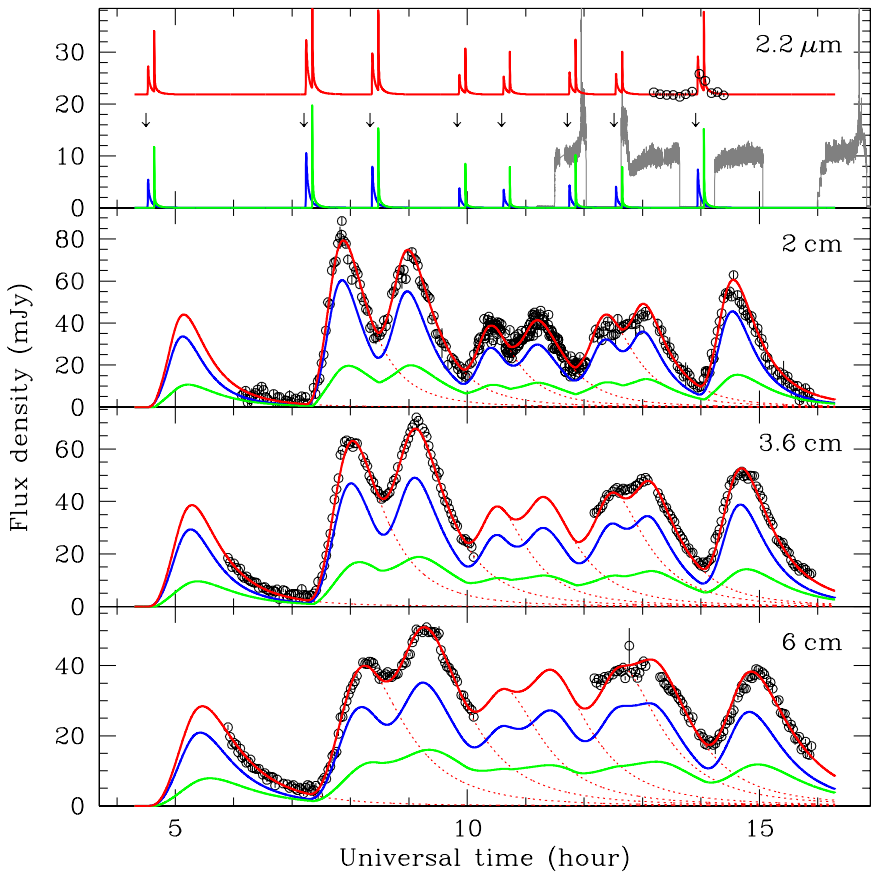}
\hfill			
\parbox[b]{55mm}{	
\caption{
	Model fit to the infrared and radio observations of GRS\,1915+105 on 15
	May 1997. The points in the four panels show the time dependence of the
	emission at observed wavelengths of 2.2\,$\mu$m, 2\,cm, 3.6\,cm and
	6\,cm, from top to bottom. The 2.2\,$\mu$m data are dereddened using a
	K-band absorption of 2.2\,mag \citep{CC04}. The continuous red line is
	the best fit model, it is the sum of the emissions from the approaching
	jet (blue line) and the receding jet (green line) plus a constant flux
	at 2.2\,$\mu$m fixed at 21.8\,mJy. Red dotted lines separate the
	emission from distinct events. In the top panel, the onset times of the
	outbursts are indicated by arrows and the grey line is the X-ray light
	curve in the 2--60\,keV band in units of 500 count\,s$^{-1}$.
}
}			
\label{fig:fit}
\end{figure*}

\section{Data and Model}
\label{sec:data}

The data we use here are the radio oscillation events of GRS\,1915+105 measured
on 15 May 1997 \citep{MDC98,DMR00}. The Very Large Array (VLA) observations
have the advantage to be very well sampled at three different radio wavelengths
(see Fig~\ref{fig:fit}). The data start with a flux decrease followed by
successive outbursts spaced by typically 1--2 hours. These oscillation events
are interpreted as ejections in a small jet that was observed simultaneously by
the Very Long Baseline Array (VLBA) \citep[see][ Fig.~9]{DMR00}. Their
evolution from short to long wavelengths is reminiscent of the observed
behaviour of the bright quasar 3C\,273 on a time scale of years \citep{TCP00}.

Additional data from the United Kingdom Infrared Telescope (UKIRT) in the
$2.2\,\mu$m infrared band show a flux enhancement which is very likely the
synchrotron precursor of the last radio outburst \citep{MDC98}. The
simultaneous X-ray measurements by the Rossi X-ray Timing Explorer (RXTE) in
the 2--60\,keV band present two flares apparently associated with two
consecutive radio outbursts. This X-ray behaviour is indeed quite different
from the light curve observed on 9 September 1997 which displays very rapid
oscillations followed by a clear X-ray dip \citep{MDC98}. A possible
interpretation of this difference is that the X-ray flares of 15 May 1997 are
not emitted by the accretion disc, but arise in the jet because of
inverse-Compton scattering of synchrotron photons.

\begin{figure}
\includegraphics[width=\hsize]{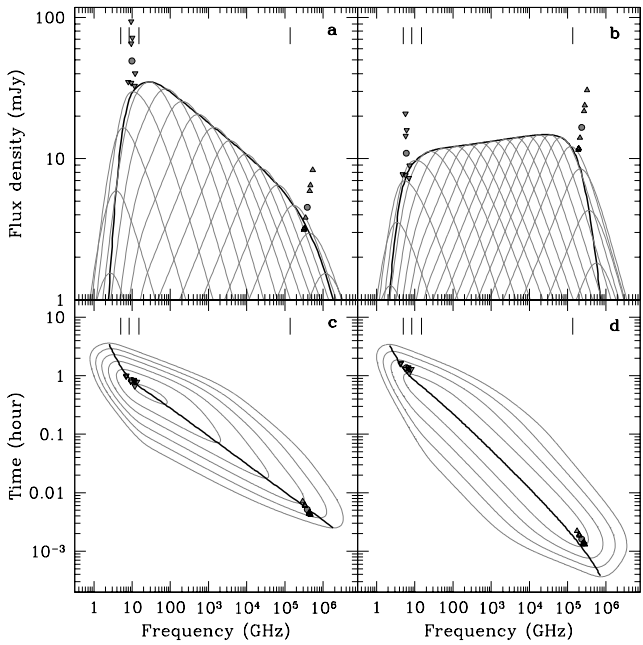}
\caption{
	Evolution of the synchrotron emission of an average model outburst in
	GRS\,1915+105 on 15 May 1997. The observed evolution is different for
	the approaching jet (\textbf{a}, \textbf{c}) and the receding jet
	(\textbf{b}, \textbf{d}). The grey lines are synchrotron spectra at
	times spaced by 0.2\,dex in \textbf{a}, \textbf{b}, and equal flux
	density contours spaced by 0.3\,dex in \textbf{c}, \textbf{d}. The
	thick black line is the path followed by the turnover of the spectrum.
	The triangles show the position of the two stage transitions for each
	outburst as compared to their average position (circles). The frequency
	of the four light curves in Fig.~\ref{fig:fit} are indicated by
	vertical lines.
}
\label{fig:evo}
\end{figure}

The \citet{MG85} shock model assumes a steady relativistic jet flow that can be
disturbed by any increase of the pressure of the jet plasma or of its bulk
velocity. If this disturbance is related to the accretion disc, it is likely to
be symmetric in both jets. Starting as a sound wave, the disturbance will
become supersonic at some point because of the decreasing pressure along the
jet. The resulting shock wave will propagate down the jet and accelerate
particles crossing the shock front. Accelerated electrons will then emit
synchrotron radiation in the increased magnetic field of the compressed plasma
behind the shock front. The evolution of the emitted spectrum depends on the
dominant energy loss mechanism of the electrons. Just after the onset of the
shock, inverse-Compton radiation is likely to be dominant, then synchrotron
losses should become more important and finally adiabatic expansion, due to the
widening of the jet, will dominate the energy losses. As a consequence, the
turnover of the self-absorbed synchrotron spectrum will move from high to low
frequencies following a characteristic three-stage path (see
Fig.~\ref{fig:evo}).

\section{Method}
\label{sec:method}

To apply the model outlined above to the observed variability of GRS\,1915+105
we use a methodology similar to the one used previously for 3C\,273
\citep{TCP00}. There are however several modifications mainly implied by the
specificity of GRS\,1915+105, but including also some more general
improvements. An important difference between the modelling of 3C\,273 and
GRS\,1915+105 is that for the latter the emission from the receding jet is well
observed \citep{MR94,FGM99,RM99} and thus cannot be neglected. We take this
into account by modelling each outburst as being the sum of twin outbursts, one
arising in the approaching jet and the other in the receding jet. The observed
emission from the two opposite jets is however different due to orientation and
relativistic effects. We model this by two parameters, one being the ratio
between the Doppler factors of the emitting material in the two jets and the
second being the observed delay between the onset of the twin outbursts. In
addition, the angle of $66\,\degr$ between the jet axis and the line of sight
\citep{FGM99} is such that the shock is likely to be viewed sideways for the
approaching jet and from behind for the receding jet \citep{MGT92}. This was
taken into account in the modelling and results in a different observed
evolution of the two jets (see Fig.~\ref{fig:evo}).

To define the specificity of each outburst, we allow to vary from one event to
the other only two out of the three parameters we used for 3C\,273
\citep{TCP00}. These two parameters are the normalization $K_{\mathrm{on}}$ of
the electron energy distribution ($N(E)=K\,E^{-s}$) and the magnetic field
strength $B_{\mathrm{on}}$ at the onset of the shock. The bulk Doppler factor
$\mathcal{D}_{\mathrm{on}}$ of the emitting material is assumed here to remain
constant. Another simplification is that there does not seem to be a constant
radio emission component in GRS\,1915+105 similar to the hot spot at the far
end of 3C\,273's jet. Furthermore, the start of the radio light curves of
GRS\,1915+105 can be simply modelled by the final decay of an outburst with
average properties.

To have a more realistic model we replaced by progressive transitions the low-
and high-frequency breaks of the synchrotron spectrum \citep[see][
Fig.~2]{TCP00} and we smoothed out the sharp edges of the three-stage
evolution. The normalization of the synchrotron spectrum was redefined as the
extrapolation of the optically thin slope, as in the original \citet{MG85}
model. Finally, we introduced a new parameter to allow the frequency ratio
$\nu_{\mathrm{h}}/\nu_{\mathrm{m}}$ of the low frequency break to the turnover
of the spectrum to vary with time. This allows to describe a synchrotron source
which is first inhomogeneous and becomes progressively homogeneous near the
transition to the final adiabatic expansion stage, as it seems to be the case
in GRS\,1915+105 (see Fig.~\ref{fig:evo}a,\,b).

The model presented here has a total of 34 free parameters. 13 parameters are
used to describe completely the shape and the evolution of the synchrotron
spectrum for an average outburst in the approaching and the receding jets. The
remaining 21 parameters are used to define the start times and the specificity
of each distinct event. This model is the simplest jet model reproducing well
the observations. It assumes a constant Doppler factor along the jet and a
linear decrease $B \propto R^{-1}$ of the magnetic field $B$ with the jet
half-width $R$, as expected if $B$ is perpendicular to the jet axis
\citep{BBR84}. The model parameters are constrained by a  simultaneous
least-square fit to the 795 radio and infrared measurements of
Fig.~\ref{fig:fit}. This is achieved by many iterative fits of small subsets of
the 34 parameters. The overall fit is shown in Fig.~\ref{fig:fit}. Although the
adjustment is not perfect (reduced $\chi^2$ of 6.4), the shape of the light
curves is very well reproduced by the model.

\section{Results and Discussion}
\label{sec:results}

The overall evolution of an average outburst in GRS\,1915+105 as derived from
the observations of 15 May 1997 is shown in Fig.~\ref{fig:evo}a,\,c for the
approaching jet and in Fig.~\ref{fig:evo}b,\,d for the receding jet. The
outburst is assumed to be intrinsically the same in both jets, but the observed
evolution is different because of orientation effects. The different viewing
angle changes the normalization, as well as the slopes of the three-stage
evolution. The first effect is due to a different relativistic Doppler factor,
while the second is due to the assumption that the approaching shock wave is
viewed sideways, while the receding one is viewed from behind.

The model parameters describing the jet properties suggest that the jet when
observed was rather dissipative and this could explain why it did not propagate
very far \citep{DMR00}. A first indication for this is the non-linear increase
of the radius $R \propto L^{\,r}$ of the jet opening with distance $L$ along
the jet. The best fit value of $r=1.52$ accounts for the rapid change of the
spectral turnover frequency with time (see Fig.~\ref{fig:evo}c,\,d). It
suggests that the inner jet of GRS\,1915+105 is more like a trumpet than a
cone. This shape could result from a pressure gradient of the material
surrounding the jet or from diverging external magnetic field lines channeling
the jet plasma. Alternatively, it might be the signature of a decelerating jet
flow. To test this hypothesis, we relax the constraint of a constant Doppler
factor along the jet. This results in a model with a slight tendency towards
deceleration, but the improvement of the fit is not significant due to the lack
of further infrared or submillimeter observations. There are however other
indications suggesting a dissipative jet flow. In particular the fact that the
obtained value of $s=2.02$ for the index of the electron energy distribution
$N(E)=K\,E^{-s}$ is close to 2. This means that there is roughly as much total
energy carried by high-energy electrons than by low-energy electrons. For
$s<2$, the strong radiative losses of the dominant high-energy electrons would
lead to a substantial pressure decrease along the jet and prevent the shock to
propagate far \citep{MG85}. That such radiative losses are not negligible is
also suggested by the slightly steeper decrease ($k=3.10$) of the normalization
$K \propto R^{-k}$ of the electron energy distribution than expected if the jet
flow was really adiabatic ($k=2(s+2)/3=2.68$).

From the ratio of the relativistic Doppler factors in the receding and the
approaching jet, we can derive the bulk speed of the emitting region. We obtain
a ratio of $0.61$, which corresponds, by assuming an angle of $66\,\degr$
between the jet axis and the line of sight \citep{FGM99}, to a speed of
$0.60\,c$, $c$ being the speed of light. This mildly relativistic value
contrasts with the speeds exceeding $0.9\,c$ derived for the giant eruptions
\citep{FGM99,RM99}, but is compatible with the hypothesis of a shock viewed
sideways in the approaching jet provided that the jet opening half-angle is of
$13\,\degr$ at least \citep{MGT92}. We can also estimate the distance from the
black hole at which the shock wave forms based on the time interval between the
onset of the outburst in the two jets. This interval is found to be $\sim
400$\,s and corresponds to a distance from the black hole measured along the
jet of $1.47\,10^{13}$\,cm, by assuming the same orientation as above. This is
roughly 1\,AU and is only about twice the separation of the binary system
\citep{GCM01}.

Concerning the specificity of individual events, represented by the points in
Fig.~\ref{fig:evo}, we note that they do mostly differ in intensity from the
average outburst. According to the model, this is due to changes of the
normalization $K_{\mathrm{on}}$ of the electron energy distribution at the
onset of the shock. The effect of significant changes of the magnetic field
strength $B_{\mathrm{on}}$ from one event to the other would lead to a clear
scatter of the points along the frequency axis, which is not observed (see
Fig.~\ref{fig:evo}a,\,b). The fact that the specificity of individual outbursts
is mainly due to differences in the normalization $K_{\mathrm{on}}$ of the
electron energy distribution is to be expected if the outbursts are due to an
increased injection rate at the base of the jet. This finding therefore
supports the idea that the shock waves forming in the jet of GRS\,1915+105 are
the result of instabilities in the inner accretion disc.

It is difficult to assess whether a scaled-up version of the shock model
presented here could apply to the giant eruptions of GRS\,1915+105. Their radio
light curve with its very fast rise compared to its slow decay \citep{DMR00}
resembles the infrared model light curve for the approaching jet, but expanded
about 1000 times in duration and 100 times in amplitude. The optically thin
synchrotron spectrum observed in the radio during the decay of major eruptions
has also roughly the same slope than the infrared spectrum deduced here. This
analogy suggests that giant eruptions could be qualitatively similar events
than those studied here, but with physical conditions such that their emission
is shifted by orders of magnitude towards lower frequencies, longer time scales
and enhanced fluxes.

\section{Conclusion}
\label{sec:conclusion}

This work shows that the microquasar GRS\,1915+105 has, at least during some
epochs, a variability behaviour similar to the quasar 3C\,273, which can be
modelled by shock waves propagating in a relativistic jet. This analogy
suggests that the physical nature of relativistic jets and their emission is
not dependent on the mass of the black hole at the origin of the ejection. The
study of jets in both quasars and microquasars is therefore complementary.
While quasars are bright and their jet structure is relatively well resolved,
microquasars have the advantage of shorter variability time scales and of a
less relativistic flow allowing us to observe also their counter-jet emission.
The ability shown here to constrain jet emission models by multi-wavelength
observations opens new perspectives for studying relativistic jets in quasars,
microquasars and tentatively even in gamma-ray bursts.

\begin{acknowledgements}
We thank V. Dhawan for kindly providing use the VLBA data.
Y.F. acknowledges financial support from the CNES.
\end{acknowledgements}

\bibliographystyle{aa}	
\bibliography{biblio} 

\end{document}